\documentclass[11pt,a4paper]{article}

\usepackage[utf8]{inputenc}
\usepackage[T1]{fontenc}
\usepackage[english]{babel}
\usepackage{lmodern}
\usepackage{geometry}
\usepackage{graphicx}
\usepackage{microtype}
\usepackage{mathtools}
\usepackage{amssymb}
\usepackage{amsthm}
\usepackage{xcolor}
\usepackage{setspace}
\usepackage[hypertexnames=false]{hyperref}
\hypersetup{colorlinks=true,linkcolor=blue,citecolor=blue,urlcolor=blue}

\newtheorem{definition}{Definition}

\title{Non-Metricity Trace Forms and Weak-Field Light Deflection:\\
A Geometric Optical Formulation}

\author{Sabutay U\u{g}ur\\[0.5em]
\normalsize Department of Physics, Pamukkale University, Denizli, Turkey\\[0.25em]
\normalsize \texttt{sugur24@posta.pau.edu.tr}}

\begin{document}

\maketitle

\begin{abstract}
This paper develops a weak-field, non-metric operational account of gravitational light deflection. In the isotropic formulation of the Schwarzschild weak field, the non-metricity trace 1-forms \(W\) and \(P\) are computed explicitly and shown to determine the optical response through the exact relation
\[
  d\ln n = W-4P,
  \qquad n=\frac{g}{f}.
\]
Thus the effective refractive index governing null propagation is not introduced as an external optical analogy, but is tied directly to the trace sector of non-metricity. To connect the trace geometry with matter, an effective scalar calibration potential \(\chi\) is sourced by an observer-projected macroscopic dilation current, yielding \(\nabla^2\chi=-\kappa\rho\) in the static weak-field limit. The point-mass calibration fixes \(\Phi=\chi/C_0\simeq2GM/(rc^2)\), and the resulting optical calculation gives the Eddington deflection angle \(4GM/(bc^2)\). The significance of the result is not a modified numerical prediction, but a closed weak-field chain from macroscopic dilation sourcing to non-metricity traces, effective refractive index, and light deflection, without postulating a microscopic deformation law for material rods or clocks.
\end{abstract}

\par

\section{Introduction}

General Relativity (GR) describes gravity through spacetime curvature, whereas Symmetric Teleparallel Gravity (STG) reformulates the same gravitational dynamics, in its GR-equivalent sector, in a curvature-free and torsion-free geometry where non-metricity carries the gravitational information \cite{Nester1999,Adak2006,BeltranJimenez2018,BeltranJimenez2020}. This reformulation raises a physical question that is not answered by formal equivalence alone: what operational role can non-metricity play in an observable weak-field phenomenon such as light deflection?

This work addresses that question by following the trace sector of non-metricity through the optical calculation. The central result is the geometric--optical bridge
\[
  d\ln n = W-4P,
  \qquad n=\frac{g}{f},
\]
where \(W\) and \(P\) are the two non-metricity trace 1-forms and \(n\) is the effective refractive index obtained from the isotropic null condition. This relation keeps the non-metric variables active in the observable chain: the weak-field optical response is expressed through non-metricity traces rather than introduced only as a curvature-based geodesic result or as a separate optical analogy.

 This point can be summarized by comparing two explanatory chains. In a purely reductive use of the GR-equivalent sector, one would proceed as
\begin{center}
\resizebox{0.95\textwidth}{!}{$
\text{STG/non-metricity formalism}
\longrightarrow
\text{GR-equivalent limit}
\longrightarrow
\text{Schwarzschild metric}
\longrightarrow
\text{light deflection}.
$}
\end{center}
In such a chain, non-metricity would function mainly as an alternative transcription of the same metric result. The present work instead keeps the non-metricity trace decomposition explicitly in the optical mechanism:
\begin{center}
\resizebox{0.95\textwidth}{!}{$
\text{STG/non-metricity formalism}
\longrightarrow
\text{trace decomposition}
\longrightarrow
W,\ P
\longrightarrow
d\ln n=W-4P
\longrightarrow
\text{optical-index gradient}
\longrightarrow
\text{light deflection}.
$}
\end{center}
This distinction concerns the mathematical explanatory chain, not a phenomenological disagreement with GR. The weak-field Eddington value is intentionally recovered. What is not handed over to the GR limit is the localization of the optical-index gradient: it is identified within the non-metricity trace sector as \(W-4P=d\ln n\).

The operational interpretation is formulated in terms of local calibration. The spatial and temporal components of the weak-field isotropic metric are read as asymmetric calibration factors relative to an asymptotic reference. This does not mean that material rods or clocks are postulated to undergo literal microscopic deformation. Rather, the known weak-field spatial--temporal asymmetry is represented operationally through the calibration structure encoded by non-metricity.

To connect the geometric trace calculation with matter, the paper introduces an effective scalar calibration potential \(\chi\). Its source equation,
\[
  \nabla^2\chi=-\kappa\rho,
\]
is obtained from a phenomenological dilation-induced scalar action. The source is the observer projection of a macroscopic dilation current, \(J=-u_\mu D^\mu\simeq\eta\rho\), motivated by the dilation sector of metric-affine gravity and hypermomentum \cite{Hehl1995,Iosifidis2020,Iosifidis2023}. This is not presented as a microscopic hypermomentum current of ordinary matter, nor as a complete affine parent theory. Its role is to supply a controlled weak-field source for the calibration potential.

With the point-mass calibration
\[
  \Phi=\frac{\chi}{C_0}\simeq\frac{2GM}{rc^2},
\]
the optical calculation reproduces the Eddington deflection angle. The purpose of this reproduction is not to compete with the GR value, but to show that the established weak-field observable can be carried through a non-metric operational chain:
\[
\text{macroscopic dilation source}
\;\longrightarrow\;
\chi
\;\longrightarrow\;
W,\ P
\;\longrightarrow\;
d\ln n
\;\longrightarrow\;
\text{light deflection}.
\]

The paper is organized as follows. Section~2 introduces the STG and non-metricity notation. Section~3 develops the operational calibration framework. Section~4 derives the effective scalar source equation from the macroscopic dilation postulate. Section~5 computes the non-metricity trace 1-forms and derives the optical bridge. Section~6 applies the resulting refractive-index structure to weak-field light deflection. Sections~7 and~8 present the discussion and conclusions.

\section{Theoretical Background}

\subsection{Symmetric Teleparallel Geometry}

In symmetric teleparallel geometry, curvature and torsion vanish while non-metricity is non-zero. This structure constitutes the geometric basis of the non-metricity-based equivalent formulation of GR \cite{Nester1999, BeltranJimenez2018, BeltranJimenez2020}. For a general classification of equivalent geometric formulations based on curvature, torsion, and non-metricity, and their modified gravity generalizations, see \cite{Heisenberg2019Review}:
\begin{equation}
  T^a = 0, \qquad R^a{}_b = 0, \qquad Q_{ab} \neq 0 .
\end{equation}
In the coincident gauge, the \emph{coordinate} connection is chosen as $\Gamma^\alpha{}_{\mu\nu} = 0$. Correspondingly, when working in a position-dependent vierbein basis, the connection 1-form arising from the basis transformation can be written as
\begin{equation}
  \omega^a{}_b = h^a{}_{\alpha}\, dh^\alpha{}_b.
\end{equation}
For coframe/differential form notation and earlier STG vierbein-based non-metricity calculations, see \cite{Nester1999, Adak2006, AdakKalaySert2006, AdakSert2013}. This expression will be used explicitly for the isotropic Schwarzschild vierbein in Section~5.

\subsection{Non-Metricity Tensor}

The non-metricity tensor is the fundamental geometric object measuring the incompatibility of the connection with the metric in metric-affine geometries \cite{Hehl1995}. In this work the sign convention is taken as
\begin{equation}
  Q_{\alpha\mu\nu}=-\nabla_\alpha g_{\mu\nu}.
\end{equation}
In the coframe/vierbein formalism, the corresponding non-metricity 1-forms are
\begin{equation}
  Q_{ab}:=\frac12(\omega_{ab}+\omega_{ba}).
\end{equation}
The two trace 1-forms used in this work are assigned distinct symbols:
\begin{equation}
  W:=\eta^{ab}Q_{ab},
  \qquad
  P:=(\iota^aQ_{ab})e^b.
  \label{eq:trace_dictionary}
\end{equation}
Here $\iota^a$ denotes contraction with the frame vector dual to $e^a$.
We denote the first trace 1-form by $W$ to avoid confusion with $Q_{ab}$ and the non-metricity scalar $\mathbb{Q}$; in the conventional differential-form notation this trace is commonly denoted by $Q$.

The disformation tensor is defined as
\begin{equation}
  L^\alpha{}_{\mu\nu}
  =\frac12 g^{\alpha\beta}
  \left(
    -Q_{\mu\beta\nu}-Q_{\nu\beta\mu}+Q_{\beta\mu\nu}
  \right),
\end{equation}
whereupon the non-metricity scalar is written as
\begin{equation}
  \mathbb{Q}
  =g^{\mu\nu}\!\left(
    L^\alpha{}_{\beta\mu}L^\beta{}_{\nu\alpha}
    -L^\alpha{}_{\beta\alpha}L^\beta{}_{\mu\nu}
  \right).
\end{equation}
For the non-metricity traces, the disformation tensor, and the modern usage of the non-metricity scalar in the symmetric teleparallel/coincident gravity context, see \cite{BeltranJimenez2018, Jimenez2020fQ}. The use of the $Q_{ab}$ 1-form structure in the coframe/vierbein formalism will be applied in Section~5 following \cite{Nester1999, AdakKalaySert2006, AdakSert2013}.

\subsection{\texorpdfstring{$f(\mathbb{Q})$}{f(Q)} Theory and Action}

In standard $f(\mathbb{Q})$ theories, the gravitational action is written as a function of the non-metricity scalar \cite{BeltranJimenez2018, Jimenez2020fQ}. For the STG/GR equivalence in the coincident gauge and in the linear limit $f(\mathbb{Q})=\mathbb{Q}$, see in particular \cite{BeltranJimenez2018}. In the normalization used in this work, the action is schematically written as
\begin{equation}
  S=\int d^4x\,\sqrt{-g}
  \left[\frac{1}{2\kappa_g}f(\mathbb{Q})+\mathcal L_m\right],
\end{equation}
where $\kappa_g=8\pi G/c^4$ is the Newton--Einstein constant. The standard $f(\mathbb{Q})$ field equations are obtained from metric and connection variations; this full derivation is not reproduced here \cite{Jimenez2020fQ}. In the coincident gauge, the connection can be set to zero by an appropriate coordinate choice.

In this work, the linear limit $f(\mathbb{Q})=\mathbb{Q}$ is recalled not as the basis of the optical calculation, but to indicate the GR-equivalent background of standard STG and the context of the GR-equivalent reduction in the literature. The source equation used below is instead obtained from the effective scalar sector established in Section~4. For this reason, unlike the standard minimal $f(\mathbb{Q})$ framework, the physical motivation invokes a metric-affine dilation sector in which matter may couple to the independent connection; for hypermomentum and the dilation current, see \cite{Hehl1995}.

To prevent the different mathematical objects from sharing the same symbol, the notation is fixed as
\begin{equation}
\begin{gathered}
  Q_{ab}\ \text{: non-metricity 1-forms},\qquad
  W=\eta^{ab}Q_{ab}\ \text{: first trace 1-form},\\
  P=(\iota^aQ_{ab})e^b\ \text{: second trace 1-form},\qquad
  \mathbb{Q}\ \text{: the scalar of }f(\mathbb{Q})\text{ gravity},\\
  C(r),\,C_0\ \text{: operational calibration quantities},\qquad
  \Phi:=\chi/C_0\ \text{: optical potential},\\
  \Psi_m\ \text{: matter fields},\qquad
  J:=-u_\mu D^\mu\ \text{: scalar dilation source}.
\end{gathered}
\label{eq:symbol_dictionary}
\end{equation}

\section{Operational Calibration Framework}

This section introduces the operational calibration picture used in the paper. The basic idea is that the presence of gravitating matter changes the representation of local spatial and temporal measurement standards relative to an asymptotically flat reference region. The calibration change is not treated as a microscopic mechanical deformation of rods or clocks. Rather, it is an effective geometrical representation of how local standards are compared with a distant reference calibration.

\subsection{Empty Space as Asymptotic Reference Calibration}

In this work, empty space is identified with the asymptotically flat region in which the local calibration standard takes its reference value. In the presence of matter, the non-metric calibration environment modifies this reference value operationally. This modification is expressed by the following phenomenological assumption:
\begin{equation}
  C(r)=C_0-\delta C(r),
  \qquad
  \delta C(r)\geq0,
  \qquad
  C(r)\leq C_0.
\end{equation}
Here $C_0$ represents the unperturbed reference calibration associated with the asymptotically flat region, and $\delta C(r)$ represents a signed local shift used to parameterize the reference-dependent distortion near a mass. The symbol $C$ is used to avoid confusion with the non-metricity 1-forms $Q_{ab}$, the trace 1-forms $W$ and $P$, and the non-metricity scalar $\mathbb{Q}$. It should not be read as the physical length of a material measuring unit.

$C_0$ is a phenomenological constant representing the dimensionless reference calibration value of the local measurement standard in empty space. Likewise, $\chi$ represents the dimensionless local calibration distortion relative to this reference value. Therefore:
\begin{equation}
  [C_0]=1,\qquad [C(r)]=1,\qquad [\chi]=1,\qquad
  \left[\frac{\chi}{C_0}\right]=1.
\end{equation}
The physically operative dimensionless optical potential is defined by
\begin{equation}
  \Phi:=\frac{\chi}{C_0}.
  \label{eq:Phi_def}
\end{equation}
At the present effective level, $C_0$ fixes the reference normalization of the calibration field $\chi$, not an independently observable new constant. The weak-field optical observables depend only on the dimensionless combination $\Phi=\chi/C_0$.

A simple analogy helps clarify the operational meaning of this calibration change. Consider a carpet of fixed length placed first on a flat floor and then on a staircase. The carpet and the measuring rod are not mechanically compressed. Nevertheless, the horizontal distance covered by the carpet on the staircase is smaller than the same carpet length on a flat floor, because part of its length is distributed along the vertical direction. In the analogy, the flat floor represents the unperturbed reference calibration associated with the asymptotically flat region, while the staircase represents a geometrically nontrivial calibration environment induced by the gravitating mass. In the same sense, the present model does not introduce a microscopic contraction law for material rods or clocks. Here ``shortening'' is only a visual appearance in the analogy; the general concept is distortion relative to a reference, not microscopic contraction. A more detailed discussion of this distinction is given in Sec.~\ref{sec:math_equiv}. This restriction is intentional: a literal matter-dependent deformation would make the measuring device itself part of the effect to be measured and would require a separate microscopic model of rods, clocks, and their coupling to the gravitational environment. Here the calibration change is instead treated as an effective geometrical representation of local spatial and temporal standards relative to a distant reference.

This structure is not derived directly from a microscopic deformation law for material measuring devices; it is adopted as an initial assumption for the operational interpretation of the light deflection phenomenon within the STG/MAG-inspired non-metric effective framework.

\subsection{Light and Local Calibration}

In empty space, since the local calibration standard remains fixed at the dimensionless reference value $C_0$ ($\delta C = 0$), the propagation path of light has a straight reference representation. In the presence of a mass distribution, the local spatial and temporal calibration components change asymmetrically. This should not be read as a physical shortening of a measuring rod. It means that local spatial and temporal intervals acquire a different projection when compared with the distant reference geometry. This asymmetry is expressed mathematically below through the relations $g(r)>1$, $f(r)<1$, and $n(r)=g(r)/f(r)$.

In STG, both photons and material measuring instruments are coupled to the same local metric. Near a mass, the spatial and temporal components of the local calibration standard change at different rates: the spatial scale factor $g(r)$ increases while the temporal scale factor $f(r)$ decreases; equivalently, $g_{ij}=g^2(r)\delta_{ij}$ grows while $|g_{00}|=f^2(r)c^2$ shrinks. This asymmetry has the following operational consequence: when the propagation path of light is represented relative to the asymptotic reference calibration, it appears curved compared to the flat reference. This operational deviation is not a geometric bending in an absolute sense; it is a measurement-representation effect arising from the asymmetric modulation of the local spatial and temporal standards.

In the following, the term ``light deflection'' denotes this operational representation effect; it refers not to a physical force acting on the photon, but to the change in the representation of its propagation path through the local calibration structure. This terminological clarification is essential for the physical interpretation proposed in this paper.

A simple carpet example can illustrate the distinction between physical deformation and reference-dependent representation. A carpet segment of length
\begin{equation}
  L_{\rm carpet}=100\,\mathrm{m}
\end{equation}
has the same physical length whether it is laid on a flat floor or over steps. If its horizontal projection on the flat reference direction is, for example,
\begin{equation}
  L_{\rm proj}=75\,\mathrm{m},
\end{equation}
this does not mean that the carpet or the measuring rod has contracted. It means that the same physical interval has a different projection relative to the chosen reference geometry.
This projection language is the intended operational meaning in the present model. The carpet example is not a model for calculating the optical index. Its only purpose is to show that the representation of an interval relative to a reference geometry can change even when the physical object itself is not deformed. Thus, the \(75\,\mathrm{m}\) horizontal projection in the analogy should not be read as implying \(n<1\). In the optical problem, the relevant quantity is fixed by the weak-field metric factors: \(g(r)>1\), \(f(r)<1\), and therefore \(n(r)=g(r)/f(r)>1\). The weak-field optical counterpart of this reference-dependent distortion is not a shortened material rod, but the spatial variation of the effective refractive index.

\subsection{Asymmetric Local Calibration: Mathematical Correspondence}

The operational description is quantified through the spatial and temporal scale factors of the isotropic Schwarzschild metric:
\begin{align}
  g(r) &\approx 1 + \frac{GM}{rc^2} > 1
    \quad \text{(spatial projection is modified relative to the reference)}, \\
  f(r) &\approx 1 - \frac{GM}{rc^2} < 1
    \quad \text{(the temporal rate slows near the mass)}.
\end{align}
This asymmetry ($g \neq f$) generates the effective refractive index
\begin{equation}
  n(r) = \frac{g(r)}{f(r)} \approx 1 + \frac{2GM}{rc^2}
  \label{eq:nref}
\end{equation}
at first order in the weak field. The effective refractive index is greater than unity ($n>1$), meaning that the optical path behaves as if the medium had a larger refractive index near the mass.

The spatial statement $g(r)>1$ should therefore be read as a statement about the metric representation of spatial intervals in the isotropic weak-field form, not as a shortening of a material measuring unit. By itself, $g(r)>1$ does not define the optical index. The effective optical index is obtained only after the spatial and temporal calibration factors are combined through $n(r)=g(r)/f(r)$. Since $g(r)>1$ and $f(r)<1$ in the weak field, this combination gives $n(r)>1$.

This observation does not introduce gravitational clock slowing or spatial scale modification as new effects; these are already encoded in the standard weak-field metric. The role of the present construction is to reinterpret this known asymmetry operationally and, in Section~5, to express it through the non-metricity trace 1-forms via the optical bridge $d\ln n=W-4P$.

The local calibration distortion parameter $\lambda(r)$ is defined as the relative shift of the local calibration value with respect to the asymptotic reference:
\begin{equation}
  \lambda(r):=\frac{C(r)-C_0}{C_0}
  =-\frac{\delta C(r)}{C_0}\leq0.
\end{equation}
The sign convention is $\lambda(r)\leq0$: near a mass the local calibration value is shifted relative to the reference. This scalar is not the optical potential itself and does not represent the full spatial--temporal asymmetry. In the weak field,
\begin{equation}
  \lambda(r)\approx-\frac{\chi(r)}{2C_0}
  =-\frac{\Phi(r)}{2}.
\end{equation}
Thus $\lambda$ is a bookkeeping parameter for the chosen negative calibration shift, whereas the physically operative optical quantity is $\Phi=\chi/C_0$, which measures the difference between the spatial and temporal factors and enters $n(r)$.


\section{Variational Derivation of the Effective Source Equation}

\subsection{Motivation}

The geometric calculation of Section~5 determines the trace 1-forms of non-metricity and their weak-field relation to the optical potential. That calculation alone does not determine a matter source equation for the scalar $\chi$. In this section a complementary effective-field-theory path is taken. An effective scalar action is introduced whose source is the observer-projected dilation current. The construction is motivated by the pure dilation sector of Metric-Affine Gravity (MAG) \cite{Hehl1995, Iosifidis2020, Iosifidis2023}.

The distinction is important: the equation obtained below follows from variation with respect to the scalar $\chi$. It is not the full affine connection equation $\delta S/\delta\Gamma^\lambda{}_{\mu\nu}=0$. The metric-affine dilation sector supplies the physical motivation and the definition of the source, while the scalar action supplies the explicit variational equation used in the static weak-field model.

\subsection{Metric-Affine Setup}

The standard metric-affine definitions and dilation-sector terminology used in this section follow \cite{Hehl1995,Iosifidis2020,Iosifidis2023}, while the coframe-based STG notation follows Adak and collaborators \cite{AdakKalaySert2006,AdakSert2013}.

Spacetime is defined by the triple $(M,g_{\mu\nu},\Gamma^\lambda{}_{\mu\nu})$, with connection independent of the metric. The symmetric teleparallel constraints are
\begin{equation}
  T^\lambda{}_{\mu\nu}=0,\qquad
  R^\lambda{}_{\rho\mu\nu}=0,\qquad
  Q_{\lambda\mu\nu}:=-\nabla_\lambda g_{\mu\nu}\neq0.
\end{equation}
The scalar $\chi$ is introduced as an effective dilation-motivated field. No identity of the form $Q_\mu=\partial_\mu\chi$ is imposed. Its weak-field relation to the geometrically calculated trace 1-forms is established only through the optical potential $\Phi=\chi/C_0$ in Section~5.

\subsection{Extended Effective Action}

In standard minimal STG, matter is usually taken to couple only to the metric. In the present construction the metric-affine dilation sector motivates the following effective scalar extension:
\begin{equation}
  S=\int d^4x\,\sqrt{-g}\left[
    \frac{1}{2\kappa_g}\mathcal L_G(g,\Gamma)
    -\frac{\beta}{2}g^{\mu\nu}\partial_\mu\chi\,\partial_\nu\chi
    +\lambda_D\chi J
    +\mathcal L_m^{(0)}(g,\Psi_m)
  \right],
  \label{eq:extaction}
\end{equation}
where $\kappa_g=8\pi G/c^4$ is the Newton--Einstein constant; $\mathcal L_G(g,\Gamma)$ represents the background torsion-free and curvature-free gravitational sector; $\beta>0$ is the kinetic coefficient of $\chi$; $\lambda_D$ is the dilation coupling constant; $J$ is the scalar dilation source; and $\mathcal L_m^{(0)}$ is the connection-independent part of the matter Lagrangian.

The $\chi$ sector is an effective scalar sector motivated by dilation hypermomentum. The geometric origin of the kinetic term from a complete affine parent action, and a variational constraint relating $\chi$ directly to a connection trace, are not assumed in the present model. Accordingly, the field equation derived below is a scalar Euler--Lagrange equation. The symbol $\lambda_D$ denotes a constant coupling and is distinct from the operational field $\lambda(r)$ introduced in Section~3.3.

The constants $\beta$, $\lambda_D$, and the macroscopic proportionality parameter introduced below are not treated as independently measurable microscopic couplings in this effective description. Their dimensions are fixed so that the action is dimensionally consistent and so that the combination $\lambda_D\eta/\beta$ has the dimensions of the weak-field source coefficient $\kappa$. After calibration to the Newtonian limit, only this combination enters the point-source solution and is fixed by $\kappa=8\pi G C_0/c^2$.

\subsection{Hypermomentum and Dilation Current}

The variation of a connection-dependent matter action with respect to the independent connection produces the hypermomentum tensor \cite{Hehl1995, Iosifidis2020}:
\begin{equation}
  \Delta^\lambda{}_{\mu\nu}
  :=-\frac{2}{\sqrt{-g}}\,
    \frac{\delta\!\left(\sqrt{-g}\mathcal L_m\right)}
         {\delta\Gamma^\lambda{}_{\mu\nu}} .
  \label{eq:hypermomentum}
\end{equation}
The hypermomentum decomposes into spin, shear, and dilation parts \cite{Hehl1995}. In this work the spin and shear components are neglected and only the trace sector is retained. The dilation current is
\begin{equation}
  D_\mu:=\Delta^\lambda{}_{\lambda\mu},
  \qquad
  D^\mu:=g^{\mu\nu}D_\nu .
  \label{eq:dilation}
\end{equation}
Iosifidis et al.\ \cite{Iosifidis2023} have studied the pure dilation regime as a distinct physical sub-sector, showing that dilation hypermomentum can produce an effective fluid contribution.

For an observer with unit four-velocity $u^\mu$, the scalar source used in the effective action is defined as
\begin{equation}
  J:=-u_\mu D^\mu .
  \label{eq:J_def}
\end{equation}
This observer projection replaces the coordinate-dependent identification $J\simeq D^0$. In the static rest frame and with the $(-,+,+,+)$ signature, the definition reduces to the corresponding dilation charge density.

\subsection{Founding Postulate: Macroscopic Dilation}

\begin{definition}[Macroscopic Dilation Postulate]
Macroscopic matter distributions carry a non-zero effective dilation charge. In the static weak-field rest frame, the observer-projected scalar dilation source is proportional to the mass density:
\begin{equation}
  J=-u_\mu D^\mu\simeq\eta\rho .
  \label{eq:postulate}
\end{equation}
Here $\eta$ is an effective coupling constant ensuring dimensional consistency.
\end{definition}

The relation \(J\simeq\eta\rho\) should be understood within the effective calibration framework introduced above. It is not introduced as a microscopic hypermomentum current of ordinary matter. In many standard macroscopic matter models, ordinary matter is taken to source the geometry only through its energy--momentum tensor, while hypermomentum is set to zero because the matter action has no independent affine-connection response. The present relation is instead a phenomenological macroscopic dilation source in the static weak-field regime. Its role is not to replace the usual energy--momentum sourcing of the metric, but to encode the effective calibration response associated with a matter distribution and to source the scalar calibration field \(\chi\). The microscopic value of $\eta$ is not separately fixed; the combination $\lambda_D\eta/\beta$ is absorbed into $\kappa$ and determined by weak-field calibration.

If this calibration response were treated as a microscopic deformation, it would become an independently observable and testable physical claim: the length of a material rod, the internal structure of a clock, or the atomic bonding scales of matter would have to change directly in a gravitational field. Such a claim is not part of the present model; if it were made, it would require a separate microscopic matter model and independent experimental evidence. In particular, the present framework does not assume that a material rod of fixed length, such as a 100 cm wooden rod defined near the Earth, undergoes an additional microscopic contraction merely by being displaced away from the Earth.

The physical plausibility of the postulate is motivated by two lines of work. First, the cosmological hyperfluid framework of Iosifidis et al.\ \cite{Iosifidis2020, Iosifidis2023} shows that the dilation component of hypermomentum can be consistently retained as an effective matter degree of freedom. Second, Ghilencea \cite{Ghilencea2024} relates mass scales and local scale breaking to non-metric geometry in Weyl and Palatini settings. These results do not derive \eqref{eq:postulate}; they motivate treating it as an effective coupling condition in the present static model.

\subsection{Derivation of the \texorpdfstring{$\chi$}{chi} Field Equation}

The $\chi$-dependent part of \eqref{eq:extaction} is
\begin{equation}
  S_\chi=\int d^4x\,\sqrt{-g}\left[
    -\frac{\beta}{2}g^{\mu\nu}\partial_\mu\chi\,\partial_\nu\chi
    +\lambda_D\chi J
  \right].
\end{equation}
In this variation $J$ is held independent of $\chi$, so $\delta J/\delta\chi=0$. Variation with respect to $\chi$ gives
\begin{equation}
  \delta S_\chi=\int d^4x\,\sqrt{-g}\left[
    -\beta g^{\mu\nu}\partial_\mu\chi\,\partial_\nu(\delta\chi)
    +\lambda_DJ\,\delta\chi
  \right].
\end{equation}
The differential operator used here is explicitly the Levi--Civita/Laplace--Beltrami operator:
\begin{equation}
  \Box_{\rm LC}\chi
  :=\frac{1}{\sqrt{-g}}
    \partial_\mu\!\left(
      \sqrt{-g}\,g^{\mu\nu}\partial_\nu\chi
    \right).
  \label{eq:boxLC}
\end{equation}
After integration by parts, including the $\sqrt{-g}$ factor, and discarding the boundary term,
\begin{equation}
  \delta S_\chi
  =\int d^4x\,\sqrt{-g}
    \left[\beta\Box_{\rm LC}\chi+\lambda_DJ\right]\delta\chi=0.
\end{equation}
Since this must vanish for arbitrary $\delta\chi$, the scalar Euler--Lagrange equation is
\begin{equation}
  \beta\Box_{\rm LC}\chi+\lambda_DJ=0 .
  \label{eq:ELeq}
\end{equation}

\subsection{Static Weak-Field Limit}

In the static regime $\partial_t\chi=0$. In the weak-curvature linearized limit, with $g_{\mu\nu}\simeq\eta_{\mu\nu}$, the operator reduces to the Laplacian:
\begin{equation}
  \Box_{\rm LC}\chi\longrightarrow\nabla^2\chi .
\end{equation}
Using the macroscopic dilation postulate \eqref{eq:postulate},
\begin{equation}
  \beta\nabla^2\chi+\lambda_D\eta\rho=0.
\end{equation}
Defining
\begin{equation}
  \kappa:=\frac{\lambda_D\eta}{\beta}>0,
  \label{eq:kappa_def}
\end{equation}
the source equation becomes
\begin{equation}
  \nabla^2\chi=-\kappa\rho .
  \label{eq:source_derivation}
\end{equation}
This equation is not an externally imposed Poisson ansatz once the effective action and the postulate are adopted; it is the weak-field scalar Euler--Lagrange equation of the proposed dilation-motivated sector. It should not be confused with the full metric-affine or \(f(\mathbb{Q})\) field equations obtained by independent variation with respect to the metric and affine connection. Its purpose is to provide a phenomenological calibration source in the weak-field regime, not to replace the complete gravitational field equations.

\subsection{\texorpdfstring{$P$}{P} Connection and Divergence Structure}

For consistency with the geometric calculation of Section~5, the weak-field optical potential is
\begin{equation}
  \Phi:=\frac{\chi}{C_0}.
\end{equation}
Section~5 gives
\begin{equation}
  P\simeq-\frac12d\Phi
  =-\frac{1}{2C_0}d\chi .
  \label{eq:P_chi}
\end{equation}
Therefore, in the static weak-field flat-background limit,
\begin{equation}
  \nabla^\mu P_\mu
  \simeq-\frac{1}{2C_0}\nabla^2\chi
  =\frac{\kappa}{2C_0}\rho>0 .
  \label{eq:div_P}
\end{equation}
This is consistent with the geometric sign structure established in Section~5.

\subsection{Point-Mass Solution and Calibration}

For a point-mass source $\rho(\mathbf x)=M\delta^{(3)}(\mathbf x)$, equation \eqref{eq:source_derivation} becomes
\begin{equation}
  \nabla^2\chi=-\kappa M\delta^{(3)}(\mathbf x).
\end{equation}
Using $\nabla^2(1/r)=-4\pi\delta^{(3)}(\mathbf x)$, the solution is
\begin{equation}
  \chi(r)=\frac{\kappa M}{4\pi r}>0 .
  \label{eq:point_solution}
\end{equation}
Through the weak-field calibration relation
\begin{equation}
  \lambda(r)\approx-\frac{\chi(r)}{2C_0},
\end{equation}
the Newton limit fixes
\begin{equation}
  \kappa=\frac{8\pi GC_0}{c^2},
  \label{eq:kappa_cal}
\end{equation}
and hence
\begin{equation}
  \frac{\lambda_D\eta}{\beta}
  =\frac{8\pi GC_0}{c^2}.
  \label{eq:calibration}
\end{equation}
At the present effective level, $C_0$ fixes the reference normalization of the calibration field $\chi$, not an independently observable new constant. The weak-field optical observables depend only on the dimensionless combination $\Phi=\chi/C_0$.

\subsection{Status of the Model}

The framework developed in this section can be classified as follows:
\begin{itemize}
  \item It is not standard minimal STG; a dilation-motivated scalar source sector is added to the STG geometric background.
  \item It is not full general MAG; spin and shear sectors are neglected, and the complete affine connection equations are not derived.
  \item It is a \emph{dilation-induced non-metric effective model}: an effective scalar framework motivated by the pure dilation component of metric-affine matter.
\end{itemize}
The source equation emerges as the field equation obtained from variation of the proposed $\chi$-sector action once the macroscopic dilation postulate is adopted. A complete affine parent action that generates the scalar kinetic term and the relation between $\chi$ and the connection traces would correspond to a stronger microscopic matter model; this lies beyond the intended operational scope of the present construction.

\clearpage

\section{Mathematical Formulation}

The isotropic Schwarzschild background and the corresponding coincident-gauge calculations of the connection and non-metricity traces used in this section also appear in Ref.~\cite{UgurAdakBagciPala2026}, where this shared geometric input is employed in the Foldy--Wouthuysen reduction of a generalized Dirac equation to investigate low-energy fermionic dynamics. In the present work, the same geometric sector is used to develop a conceptual and operational interpretation of non-metricity through its trace structure and associated optical response.

\subsection{Vierbein and Connection 1-Form}

We use isotropic coordinates
\begin{equation}
  x^\mu=(x^0,x,y,z),\qquad x^0=ct,\qquad
  r=(x^2+y^2+z^2)^{1/2},
\end{equation}
with
\begin{equation}
  ds^2=-f^2(r)(dx^0)^2+g^2(r)(dx^2+dy^2+dz^2).
  \label{eq:isotropic_metric}
\end{equation}
For the isotropic Schwarzschild geometry,
\begin{equation}
  f=\frac{1+\xi}{1-\xi},
  \qquad
  g=(1-\xi)^2,
  \qquad
  \xi=-\frac{GM}{2rc^2}.
  \label{eq:isotropic_fg}
\end{equation}
The orthonormal coframe is chosen as
\begin{equation}
  e^0=f\,dx^0,
  \qquad
  e^1=g\,dx,
  \qquad
  e^2=g\,dy,
  \qquad
  e^3=g\,dz,
\end{equation}
so that
\begin{equation}
  h^a{}_\alpha=\operatorname{diag}(f,g,g,g),
  \qquad
  h^\alpha{}_b=\operatorname{diag}(f^{-1},g^{-1},g^{-1},g^{-1}).
\end{equation}
In the coincident gauge, the coordinate connection vanishes, while the position-dependent orthonormal basis carries the connection 1-form
\begin{equation}
  \omega^a{}_b=h^a{}_\alpha\,dh^\alpha{}_b .
\end{equation}
Since
\begin{equation}
  dr=\frac{x\,dx+y\,dy+z\,dz}{r}
  =\frac{x e^1+y e^2+z e^3}{rg},
\end{equation}
the nonzero diagonal components are
\begin{equation}
  \omega^0{}_0=-\frac{f'}{f}\,dr,
  \qquad
  \omega^1{}_1=\omega^2{}_2=\omega^3{}_3
  =-\frac{g'}{g}\,dr ,
  \label{eq:omega_diag}
\end{equation}
where a prime denotes $d/dr$. This pure-gauge connection satisfies
\begin{equation}
  T^a=0,\qquad R^a{}_b=0 .
\end{equation}

\subsection{Non-Metricity 1-Forms and Their Traces}

With signature $\eta_{ab}=\operatorname{diag}(-1,1,1,1)$ and symmetrization
\begin{equation}
  A_{(ab)}:=\frac12(A_{ab}+A_{ba}),
\end{equation}
the non-metricity 1-forms are
\begin{equation}
  Q_{ab}:=\frac12(\omega_{ab}+\omega_{ba})
  =\omega_{(ab)} .
\end{equation}
Lowering the first Lorentz index in \eqref{eq:omega_diag} gives
\begin{equation}
  Q_{00}=+\frac{f'}{f}\,dr,
  \qquad
  Q_{11}=Q_{22}=Q_{33}
  =-\frac{g'}{g}\,dr ,
  \label{eq:Qab_diag}
\end{equation}
with all off-diagonal components zero.

The first trace 1-form is
\begin{align}
  W
  :=\eta^{ab}Q_{ab}
  &=Q^{a}{}_{a}\nonumber\\
  &=Q^{0}{}_{0}+Q^{1}{}_{1}+Q^{2}{}_{2}+Q^{3}{}_{3}\nonumber\\
  &=-\left(\frac{f'}{f}+3\frac{g'}{g}\right)dr .
  \label{eq:W_exact}
\end{align}
For the \((-+++) \) signature,
\[
  Q^{0}{}_{0}=-Q_{00},
  \qquad
  Q_{00}=-Q^{0}{}_{0},
  \qquad
  Q^{i}{}_{i}=Q_{ii}\quad (i=1,2,3).
\]
The second trace 1-form is
\begin{align}
  P
  :=(\iota^aQ_{ab})e^b
  =-\frac{g'}{g}\,dr .
  \label{eq:P_exact}
\end{align}
These are distinct 1-forms; neither is the scalar $\mathbb Q$ of $f(\mathbb Q)$ gravity.

\subsection{Geometric--Optical Bridge}

The effective refractive index of the isotropic metric is
\begin{equation}
  n(r)=\frac{g(r)}{f(r)} .
\end{equation}
Using \eqref{eq:W_exact} and \eqref{eq:P_exact},
\begin{align}
  W-4P
  &=-\left(\frac{f'}{f}+3\frac{g'}{g}\right)dr
    +4\frac{g'}{g}dr\nonumber\\
  &=\left(\frac{g'}{g}-\frac{f'}{f}\right)dr
  =d\ln\!\left(\frac{g}{f}\right).
\end{align}
Hence the exact bridge is
\begin{equation}
  \boxed{d\ln n=W-4P}.
  \label{eq:geometric_optical_bridge}
\end{equation}

Within the effective optical description adopted here, light does not respond to the two non-metricity trace 1-forms separately; rather, the relevant optical gradient is encoded in their specific combination, $W-4P=d\ln n$.

In the weak field, with $m:=GM/c^2$,
\begin{equation}
  f\simeq1-\frac{m}{r},
  \qquad
  g\simeq1+\frac{m}{r},
\end{equation}
and therefore
\begin{equation}
  \frac{f'}{f}\simeq+\frac{m}{r^2},
  \qquad
  \frac{g'}{g}\simeq-\frac{m}{r^2}.
\end{equation}
Equations \eqref{eq:W_exact} and \eqref{eq:P_exact} reduce to
\begin{equation}
  W\simeq\frac{2GM}{r^2c^2}\,dr,
  \qquad
  P\simeq\frac{GM}{r^2c^2}\,dr,
\end{equation}
so that
\begin{equation}
  \boxed{P\simeq\frac12W}
  \qquad\text{(weak field)}.
  \label{eq:P_half_W}
\end{equation}

The calibrated optical potential is
\begin{equation}
  \Phi:=\frac{\chi}{C_0}
  \simeq\frac{2GM}{rc^2}.
\end{equation}
Consequently,
\begin{equation}
  d\Phi=-\frac{2GM}{r^2c^2}\,dr,
\end{equation}
and the weak-field sign chain is
\begin{equation}
  \boxed{
    W\simeq-d\Phi,
    \qquad
    P\simeq\frac12W\simeq-\frac12d\Phi
  }.
  \label{eq:weak_sign_chain}
\end{equation}
The minus sign therefore relates the geometric trace 1-forms to the radially decreasing optical potential; it does not occur between $P$ and $W$.

\subsection{Consistency with the Effective Source Equation}

Because $\Phi=\chi/C_0$, equation \eqref{eq:weak_sign_chain} gives
\begin{equation}
  P\simeq-\frac{1}{2C_0}d\chi .
\end{equation}
Taking the weak-field divergence and using \eqref{eq:source_derivation},
\begin{equation}
  \nabla^\mu P_\mu
  \simeq-\frac{1}{2C_0}\nabla^2\chi
  =\frac{\kappa}{2C_0}\rho>0 .
\end{equation}
Thus the geometric trace calculation, the decreasing radial optical potential, and the effective scalar source equation have mutually consistent signs.

\section{Null Geodesic and the Eddington Result}

\subsection{Effective Refractive Index}

In terms of the potential-like scalar field $\chi(r)$, the effective refractive index is derived from the null condition in the isotropic metric ($d\ell/dt = cf/g$):
\begin{equation}
  n(r) = \frac{g(r)}{f(r)} \approx 1 + \frac{\chi(r)}{C_0}
  \qquad \text{(weak field, first order)}
\end{equation}
This expression is consistent with \eqref{eq:nref} in Section~3.3; using $\chi(r)=\kappa M/(4\pi r)$ and $\kappa = 8\pi GC_0/c^2$ here yields an optical response consistent with the standard Newton limit. Dimensionally, since $C_0$ and $\chi$ are dimensionless calibration quantities (see Section~3.1), the ratio $\chi/C_0$ is interpreted as the dimensionless optical potential directly contributing to the refractive index.

Although this expression may appear to be merely an abstract scalar, the content of the $\chi$ potential directly reflects the asymmetry between the spatial and temporal components of the isotropic Schwarzschild metric. In the weak field, $f \approx 1 - GM/rc^2$ and $g \approx 1 + GM/rc^2$, so:
\begin{equation}
  \frac{\chi(r)}{C_0} = \frac{2GM}{rc^2} \approx g(r) - f(r).
\end{equation}
Thus $\chi/C_0$ encodes the difference between the spatial ($g$) and temporal ($f$) components of the metric. Consequently, the effective refractive index $n(r) \approx 1 + (g-f)$ contains the $g_{00} \neq g_{ij}$ asymmetry that produces relativistic light deflection. Within this framework, the bridge between the operational analogy in Section~3 and the mathematical structure here can be constructed as follows: the change in the local calibration standard reflects, in physical content, this metric asymmetry at the operational level.

\subsection{Light Path Equation}

Null condition $ds^2 = 0$ in the isotropic metric:
\begin{equation}
  -f^2 c^2\, dt^2 + g^2(dx^2+dy^2+dz^2) = 0.
\end{equation}
Here $b$ is defined geometrically as the closest-approach distance (impact parameter) of the unperturbed straight light path to the center of mass at zeroth order. $\ell$ is the path parameter taken along the straight light path at zeroth order, identified with the coordinate length in the flat background geometry. Defining the dimensionless optical potential as
\[
  \Phi(r) := \frac{\chi(r)}{C_0},
\]
the light path equation in magnitude form is
\begin{equation}
  \left|\frac{d^2 \mathbf{x}_\perp}{d\ell^2}\right|
  = \left|\nabla_\perp \ln n(r)\right|
  \approx \left|\nabla_\perp \Phi\right|
  = \frac{|\nabla_\perp \chi|}{C_0}.
\end{equation}
The magnitude notation is used here; to avoid dependence of the sign on the coordinate system, the deflection angle is expressed in terms of $|\alpha|$.

\subsection{Deflection Angle}

\begin{equation}
  |\alpha|
  = \left|
  \int_{-\infty}^{+\infty}
  \nabla_\perp \ln n\, d\ell
  \right|
  \approx
  \left|
  \int_{-\infty}^{+\infty}
  \nabla_\perp \Phi\, d\ell
  \right|
  = \frac{1}{C_0}
  \left|
  \int_{-\infty}^{+\infty}
  \frac{\partial\chi}{\partial x_\perp}\,d\ell
  \right|
\end{equation}

For $\chi = \kappa M / 4\pi r$, under the approximation that the light path is straight at zeroth order and the deflection is computed at first order under the condition $GM/bc^2 \ll 1$:
\begin{equation}
  |\alpha|
  = \frac{\kappa M}{4\pi C_0} \cdot \frac{2}{b}
\end{equation}

Substituting \eqref{eq:kappa_cal}:
\begin{equation}
  |\alpha| = \frac{4GM}{bc^2}
  \label{eq:eddington}
\end{equation}

Here $|\alpha|$ denotes the magnitude of the deflection angle.

For $M=M_\odot$ and $b = R_\odot$, $|\alpha| = 1.75$ arcseconds. This is consistent with Eddington's 1919 solar eclipse observation \cite{Eddington1920}.

\section{Discussion}

The main result of the paper is the trace-level optical bridge
\[
  d\ln n=W-4P,
  \qquad n=\frac{g}{f}.
\]
This identity gives a direct operational role to the two non-metricity trace 1-forms in the weak-field optical response. The effective refractive index is therefore not appended to the geometry as an external analogy; it is recovered from the same non-metric structure that carries the symmetric teleparallel description.

The scalar and geometric parts of the construction play complementary roles. The non-metricity traces \(W\) and \(P\) are computed directly from the isotropic vierbein in the coincident-gauge formulation. The scalar calibration potential \(\chi\), on the other hand, supplies the matter-sourced weak-field normalization through
\[
  \nabla^2\chi=-\kappa\rho,
  \qquad
  \Phi=\frac{\chi}{C_0}.
\]
The point-mass solution fixes \(\Phi\simeq2GM/(rc^2)\), so the optical bridge leads to the standard first-order refractive-index structure \(n\simeq1+\Phi\).

The agreement with the Eddington angle is a consistency requirement, not the sole content of the work. A viable weak-field model must reproduce
\[
  |\alpha|=\frac{4GM}{bc^2}
\]
in the solar weak-field regime. The contribution here is that the result is reached through an explicit non-metric chain: macroscopic dilation sourcing, scalar calibration potential, trace 1-forms, effective refractive index, and optical deflection. In this sense, the calculation does more than transfer a GR observable into different notation; it identifies the specific trace combination \(W-4P\) that carries the optical gradient.

The construction remains an effective weak-field model. It is not standard minimal STG, because it introduces an additional dilation-motivated scalar calibration sector. It is also not full metric-affine gravity, because the spin and shear sectors are neglected and the complete affine connection equations are not derived. The source \(J=-u_\mu D^\mu\simeq\eta\rho\) should therefore be read as a macroscopic effective dilation source, not as a microscopic hypermomentum current of ordinary matter. This restriction is essential: a microscopic interpretation would require a matter model capable of predicting testable changes in rods, clocks, or atomic scales, which is not part of the present framework.

The normalization \(C_0\) fixes the reference scale of the effective calibration potential. At the present level, observables depend on the dimensionless ratio \(\Phi=\chi/C_0\), while the coupling
\[
  \kappa=\frac{8\pi GC_0}{c^2}
  =\frac{\lambda_D\eta}{\beta}
\]
sets the weak-field calibration. A more fundamental affine parent action could in principle determine the scalar sector and the status of \(C_0\) from first principles, but that stronger construction is beyond the scope of the present paper.

\subsection{Mathematical Equivalence and Physical Explanation}
\label{sec:math_equiv}

The GR-equivalent sector of STG can reproduce the same weak-field observables as GR. This mathematical equivalence is necessary but does not exhaust the physical interpretation of the non-metric variables. If the calculation is closed merely by invoking GR equivalence, the non-metricity tensor becomes a transcription device rather than an active explanatory object.

The present construction keeps the non-metric trace 1-forms in the phenomenological chain. The observable optical response is attached to the exact combination
\[
  W-4P=d\ln n,
\]
and the matter dependence is encoded through the effective scalar calibration potential. The resulting interpretation is therefore not based on a new metric or a different light-deflection law, but on assigning operational content to the non-metric trace sector in the weak-field regime.

The distinction is central to the interpretation of the result. In the reductive route,
\begin{center}
\resizebox{0.95\textwidth}{!}{$
\text{STG/non-metricity formalism}
\longrightarrow
\text{GR-equivalent limit}
\longrightarrow
\text{Schwarzschild metric}
\longrightarrow
\text{light deflection}.
$}
\end{center}
the non-metric variables do not acquire an independent operational role in the phenomenon. In the trace-based route developed here,
\begin{center}
\resizebox{0.95\textwidth}{!}{$
\text{STG/non-metricity formalism}
\longrightarrow
\text{trace decomposition}
\longrightarrow
W,\ P
\longrightarrow
d\ln n=W-4P
\longrightarrow
\text{optical-index gradient}
\longrightarrow
\text{light deflection}.
$}
\end{center}
the optical gradient itself is assigned to a definite non-metricity trace combination. This is the sense in which the construction goes beyond merely taking the GR limit: it does not change the Eddington value, but it changes where the weak-field optical response is geometrically located.

This is the sense in which non-metricity is given a physical interpretation in the present model: it describes distortion relative to a reference geometry, not microscopic deformation of material objects.

The word ``shortening'' in the carpet analogy should therefore not be understood as a literal physical contraction. It describes only one possible appearance of a more general effect: a local interval may have a different projection when compared with a reference geometry. Such a change may look like shortening in the carpet example, but in optical propagation it appears as a bending of the light path. In this sense, the relevant idea is not microscopic contraction, but geometric distortion. In the present model, this distortion is measured in the weak-field optical regime by the variation of the effective refractive index, and this variation is given by
\[
  d\ln n=W-4P.
\]

The limitations are clear. The macroscopic dilation postulate is not derived from a microscopic hyperfluid model; the scalar sector is not obtained from a complete affine parent action; strong-field behavior is not addressed; and independent weak-field observables, such as Shapiro delay, are left for future analysis. These limitations do not affect the internal weak-field result, but they define the domain in which the proposed interpretation should be read.

\section{Conclusion}

This work has formulated a weak-field operational interpretation of non-metricity in which the optical response of light is governed by the trace sector of the non-metricity 1-forms. For the isotropic Schwarzschild weak field, the first and second traces
\[
  W=\eta^{ab}Q_{ab},
  \qquad
  P=(\iota^aQ_{ab})e^b
\]
satisfy the exact relation
\[
  d\ln n=W-4P,
  \qquad n=\frac{g}{f}.
\]
This equation is the central result of the paper: it identifies the non-metric trace combination that directly generates the gradient of the effective refractive index.

The matter dependence is introduced through an effective scalar calibration potential \(\chi\) sourced by the observer-projected macroscopic dilation current,
\[
  J=-u_\mu D^\mu\simeq\eta\rho.
\]
Variation of the effective scalar action gives
\[
  \nabla^2\chi=-\kappa\rho,
\]
and the point-mass calibration yields
\[
  \Phi=\frac{\chi}{C_0}\simeq\frac{2GM}{rc^2},
  \qquad
  \kappa=\frac{8\pi GC_0}{c^2}.
\]
In the weak field,
\[
  W\simeq-d\Phi,
  \qquad
  P\simeq-\frac12d\Phi,
\]
so the scalar calibration potential, the non-metric trace calculation, and the optical index are mutually consistent.

Using \(n\simeq1+\Phi\), the null-condition/effective-optical calculation gives
\[
  |\alpha|=\frac{4GM}{bc^2},
\]
the Eddington weak-field light-deflection angle. The result is not presented as a modified prediction. Its significance is that the established deflection can be obtained through a closed non-metric operational chain rather than by stopping at GR equivalence. The framework therefore assigns a concrete weak-field role to non-metricity traces while remaining explicitly effective: it does not postulate microscopic deformation of rods or clocks, and it does not claim to provide the full affine parent theory. Extending the scalar sector to a complete metric-affine derivation and testing the same trace-based interpretation against additional weak-field observables remain natural next steps.

\section*{Acknowledgement}

 The metric assumption used as part of the geometrical background of this work was adopted, and the coincident-gauge computation of the non-metricity tensor and its trace 1-forms was carried out, together with Prof. Dr. Muzaffer Adak during the author’s master’s thesis at Pamukkale University. The author sincerely thanks Prof. Dr. Adak for his guidance and collaboration during that thesis work.

\section*{Statements and Declarations}

\subsection*{Funding}

The author received no funding for this work.

\subsection*{Competing Interests}

The author declares no competing interests.

\subsection*{Data Availability}

No data were used or generated during the current study.



\begin{thebibliography}{99}

\bibitem{Nester1999}
J.~M. Nester and H.-J. Yo,
\textit{Symmetric teleparallel general relativity},
Chin. J. Phys. \textbf{37}, 113--117 (1999),
arXiv:gr-qc/9809049.

\bibitem{BeltranJimenez2018}
J.~Beltr\'{a}n Jim\'{e}nez, L.~Heisenberg and T.~Koivisto,
\textit{Coincident general relativity},
Phys. Rev. D \textbf{98}, 044048 (2018),
arXiv:1710.03116 [gr-qc].

\bibitem{BeltranJimenez2020}
J.~Beltr\'{a}n Jim\'{e}nez, L.~Heisenberg and T.~Koivisto,
\textit{The geometrical trinity of gravity},
Universe \textbf{5}, 173 (2019),
arXiv:1903.06830 [gr-qc].

\bibitem{Heisenberg2019Review}
L.~Heisenberg,
\textit{A systematic approach to generalisations of General Relativity and their cosmological implications},
Phys. Rep. \textbf{796}, 1--113 (2019).

\bibitem{Jimenez2020fQ}
J.~Beltr\'{a}n Jim\'{e}nez, L.~Heisenberg, T.~Koivisto and S.~Pekar,
\textit{Cosmology in $f(Q)$ geometry},
Phys. Rev. D \textbf{101}, 103507 (2020),
arXiv:1906.10027 [gr-qc].

\bibitem{Adak2006}
M.~Adak,
\textit{The Symmetric Teleparallel Gravity},
Turk. J. Phys. \textbf{30}, 379--390 (2006).

\bibitem{Eddington1920}
F.~W. Dyson, A.~S. Eddington and C.~Davidson,
\textit{A determination of the deflection of light by the Sun's
gravitational field, from observations made at the total eclipse
of May 29, 1919},
Phil. Trans. R. Soc. A \textbf{220}, 291--333 (1920).

\bibitem{AdakKalaySert2006}
M.~Adak, M.~Kalay and \"{O}.~Sert,
\textit{Lagrange formulation of the symmetric teleparallel gravity},
Int. J. Mod. Phys. D \textbf{15}, 619--634 (2006),
arXiv:gr-qc/0505025.

\bibitem{AdakSert2013}
M.~Adak, \"{O}.~Sert, M.~Kalay and M.~Sar\i,
\textit{Symmetric teleparallel gravity: Some exact solutions and spinor couplings},
Int. J. Mod. Phys. A \textbf{28}, 1350167 (2013),
arXiv:0810.2388 [gr-qc].

\bibitem{UgurAdakBagciPala2026}
S.~Ugur, M.~Adak, A.~Bagci and C.~Pala,
\textit{Foldy--Wouthuysen transformation of the generalized Dirac equation in symmetric teleparallel gravity},
arXiv:2607.27889 [gr-qc] (2026).

\bibitem{Hehl1995}
F.~W. Hehl, J.~D. McCrea, E.~W. Mielke and Y.~Ne'eman,
\textit{Metric-affine gauge theory of gravity: field equations,
Noether identities, world spinors, and breaking of dilation invariance},
Phys. Rep. \textbf{258}, 1--171 (1995).

\bibitem{Iosifidis2020}
D.~Iosifidis,
\textit{Cosmological hyperfluids, torsion and non-metricity},
Eur. Phys. J. C \textbf{80}, 1042 (2020).

\bibitem{Iosifidis2023}
D.~Iosifidis, I.~G. Vogiatzis and C.~G. Tsagas,
\textit{Friedmann-like universes with non-metricity},
Eur. Phys. J. C \textbf{83}, 216 (2023).

\bibitem{Ghilencea2024}
D.~M. Ghilencea,
\textit{Non-metric geometry as the origin of mass in gauge theories of scale invariance},
Eur. Phys. J. C \textbf{83}, 176 (2023),
arXiv:2203.05381 [hep-th].


\end{thebibliography}
\end{document}